\documentclass[a4paper,10pt]{scrartcl}

\usepackage{amsmath}
\usepackage{amsfonts}
\usepackage{amssymb}
\usepackage{graphicx}
\usepackage[english]{babel}
\usepackage{wasysym}
\usepackage{units}

\newcommand{\ad}{\dot{a}}

\newcommand{\qq}{\qquad}
\newcommand{\q}{\quad}
\newcommand{\reff}[1]{(\ref{#1})}
\newcommand{\vs}[1]{\vspace{#1mm}}
\newcommand{\vsO}{\vspace{.1cm}\hfill\\}
\newcommand{\vsT}{\vspace{.2cm}\hfill\\}

\newcommand{\pr}{{\scriptscriptstyle\prime}}
\newcommand{\primo}{^{\scriptscriptstyle\prime}}

\newcommand{\ao}{\bar{a}}
\newcommand{\rhoo}{\rho_{\scriptscriptstyle 0}}
\newcommand{\azero}{a_{\scriptscriptstyle 0}}
\newcommand{\so}{s_{\scriptscriptstyle 0}}
\newcommand{\tzero}{t_{\scriptscriptstyle 0}}
\newcommand{\rs}{r^S_\odot}
\newcommand{\Ls}{L^*_\odot}
\newcommand{\Lss}{L^{**}_\odot}
\newcommand{\Hzero}{H_{\scriptscriptstyle 0}}
\newcommand{\Ss}{S}
\newcommand{\Split}[2]{\begin{minipage}[c]{0.4\textwidth}
#1
\end{minipage}\begin{minipage}[c]{0.6\textwidth}
#2
\end{minipage}
}
\newcommand{\Splitt}[2]{\begin{minipage}[c]{0.6\textwidth}
#1
\end{minipage}\begin{minipage}[c]{0.4\textwidth}
#2
\end{minipage}
}

\title{\large ON THE VIABILITY OF A NON-ANALYTICAL f(R)-THEORY}

\author{\normalsize Nakia Carlevaro$^{\;a}$, Giovanni Montani$^{\;a,\;b,\;d}$
and Massimiliano Lattanzi$^{\;a,\;c}$\vsT\vsO
\emph{\footnotesize $^a$ Department of Physics - ``Sapienza'' University of Rome}\vs{-2.5}\\
\emph{\footnotesize c/o Dip. Fisica - ``Sapienza'' Universit\`a di Roma, P.le A. Moro, 5 (00185), Roma (Italia)}\\
\emph{\footnotesize $^b$ ENEA -- C.R. Frascati (Rome), UTFUS-MAG}\\
\emph{\footnotesize $^c$ ICRA -- International Center for Relativistic Astrophysics}\\
\emph{\footnotesize $^{d}$ ICRANet -- International Center for Relativistic Astrophysics Network}\\
\vsO
{\footnotesize\ttfamily nakia.carlevaro@gmail.com\quad\qquad giovanni.montani@frascati.enea.it\qq\q lattanzi@icra.it}
}
\date{}
\begin{document}
\maketitle

%
\hrule
\begin{abstract} \textbf{Abstract:} In this paper, we show how a \emph{power-law} correction to the Einstein-Hilbert action provides a viable modified theory of gravity, passing the Solar-System tests, when the exponent is between the values 2 and 3. Then, we implement this paradigm on a cosmological setting outlining how the main phases of the Universe thermal history are properly reproduced.

As a result, we find two distinct constraints on the characteristic length scale of the model, \emph{i.e.}, a lower bound from the Solar-System test and an upper one by guaranteeing the matter dominated Universe evolution.

\vsO \emph{PACS}: 95.30.Wi, 51.20.+d
\end{abstract}
\hrule

\section{Basic statements}

From the very beginning, the possibility to riformulate General Relativity by using a generic function of the Ricci scalar (see, for example, \cite{revi} for a recent review and references therein) appeared as a natural issue offered by the fundamental principles established by Einstein. However, it is important to remark that any modification of the Einstein-Hilbert (EH) Lagrangian is reflected onto a deformed gravitational-field dynamics at any length scale investigated or observed. Thus, the success of such $f(R)$ gravity in the solution of a specific problem has to match consistency with observation in different length scales \cite{zac, will, tak}. A viable self-consistent model can be often obtained at the price to consider a generalized gravitational lagrangian containing a large number of free parameters. Nevertheless, the wide spectrum of possible choices for $f(R)$ can appear as a weakness point in view of the predictivity of the theory, because a significant degree of degeneracy is expected in the model.

Here, we consider an opposite point of view, by studying the viability of a power-law correction to the EH action having a single free parameter (a length scale) once the power-law exponent is fixed. We investigate the implementation of the Solar-System test to our model \cite{lecian-solar} and then we pursue a cosmological study of the resulting modified Friedmann-Lema\^itre-Robertson-Walker (FLRW) dynamics. As expected, this scenario gives us a rather stringent range of variation for the free length scale where searching for new gravitational physics.

\section{Non-analytical power-law f(R) model}
In this paper, we consider the following modified gravitational action in the so-called \emph{Jordan frame}
\begin{equation}\label{nonan}
\Ss=-\tfrac{1}{2\chi}\;\textstyle{\int}d^{4}x\,\sqrt{-g}\;f(R)\;,\qq\q
f(R)=R+qR^n\;,
\end{equation}
where $n$ is a non-integer dimensionless parameter and $q<0$ has dimensions of $[L]^{2n-2}$ (in the equation above $\chi=8\pi G$, using $c=1$ and $G$ being the Newton constant, moreover, the signature is set as $[\,+,-,-,-\,]$). Such a form of $f(R)$ gives the following constraints for $n$: if $R>0$, all $n$-values are allowed; if $R<0$, the condition $n=\ell/(2m+1)$
must hold (where, here and in the following, $m$ and $\ell$ denote positive integer). It is straightforward to verify that $S$ in eq.\reff{nonan} is non-analytical in $R=0$ for non-integer, rational $n$, \emph{i.e.}, it does not admit Taylor expansion near $R=0$.

Let us now define the \emph{characteristic length scale} of our model as 
\begin{align}\label{cara-lenght}
L_q(n)\equiv|q|^{1/(2n-2)}\;,
\end{align}
while variations of the total action $\Ss_{tot}=\Ss+\Ss_M$ (where $\Ss_M$ denotes the matter term) with respect to the metric give, after manipulations and modulo surface terms: 
\begin{align}\label{variational}
f\primo R_{\mu\nu}-\tfrac{1}{2}\,g_{\mu\nu}f-\nabla_\mu\nabla_\nu\,f\primo +g_{\mu\nu}\Box\,f\primo =
\chi\,T_{\mu\nu}\;,
\end{align}
where $T_{\mu\nu}$ is the Energy-Momentum Tensor (EMT). Here and in the following  $(...)\primo$ indicates the derivative with respect to $R$, $\Box\equiv g^{\rho\sigma}\nabla_\rho\nabla_\sigma$ and $\nabla_\mu$ or $(...)_{;}$ denotes the covariant derivative (Greek indices run form $0$ to $3$).

We can gain further information on the value of $n$ by analyzing the conditions that allow for a consistent weak-field stationary limit. Having in mind to investigate the weak field limit of our theory to obtain predictions at Solar-System scales, we can decompose the corresponding metric as $g_{\mu\nu}=\eta_{\mu\nu}+ h_{\mu\nu}$, where $h_{\mu\nu}$ is a small (for our case, static) perturbation of the Minkowskian metric $\eta_{\mu\nu}$. In this limit, the vacuum Einstein equations read
\begin{subequations}
\end {subequations}
\begin{align}\label{mfe1}
R_{\mu\nu}-\tfrac{1}{2}\eta_{\mu\nu}R-nq (R^{n-1})_{;\mu;\nu}+nq\eta_{\mu\nu}\Box R^{n-1}=0\;,
\qq\qq\q R=3nq\Box R^{n-1}\;.
\end{align}
The structure of such field equations leads us to focus our attention on the restricted region of the parameter space $2<n<3$. This choice is enforced by the fulfillment of the conditions by which all other terms are negligible with respect to the linear and the lowest-order non-Einsteinian ones.

\section{Viability of the theory: the Solar-System test}

From the analysis of the weak-field limit in the Jordan frame, \emph{i.e.}, eqs.\reff{mfe1}, we learn the possibility to find a post-Newtonian solution by solving eqs.\reff{mfe1} up to the next-to-leading order in $h$, \emph{i.e.}, up to $\mathcal{O}(h^{n-1})$, and neglecting the $\mathcal{O}(h^2)$ contribution only for the cases $2<n<3$. These considerations motivate the choice we claimed above concerning the restriction of the parameter $n$. 

The most general spherically-symmetric line element in the weak-field limit is
\begin{equation}\label{line}
ds^2=(1+\Phi)dt^2-(1-\Psi)dr^2-r^{2}d\Omega^2\;,
\end{equation}
where $\Phi$ and $\Psi$ are the two generalized gravitational potentials and $d\Omega^2$ is the solid angle element. Within this framework, the modified Einstein equations \reff{mfe1} rewrite
\vspace{0mm}\\
\Splitt{
\begin{align}
R_{tt}-\tfrac{1}{2}R-nq\nabla^2 R^{n-1}=0\;,\nonumber\\
R_{rr}+\tfrac{1}{2}R-nq (R^{n-1})_{,r,r}+nq\nabla^2 R^{n-1}=0\;,\nonumber\\
R_{\theta\theta}+\tfrac{1}{2}r^2R-nq r(R^{n-1})_{,r}+nq r^2\nabla^2 R^{n-1}=0\;,\nonumber\\
R+3nq\nabla^2 R^{n-1}=0\;,\nonumber\\
\left[ \nabla^2\equiv \tfrac{d^2}{dr^2}+\tfrac{2}{r}\tfrac{d}{dr}\right]\;,\nonumber
\end{align}
}{
\begin{subequations}
\begin{align}
&R=\nabla^2\Phi+\tfrac{2}{r^2}(r\Psi)_{,r}\;,\nonumber\\
&R_{tt}=\tfrac{1}{2}\nabla^2\Phi\;,\nonumber\\
&R_{rr}=-\tfrac{1}{2}\Phi_{,r,r}-\tfrac{1}{r}\Psi_{,r}\;,\nonumber\\
&R_{\theta\theta}=-\Psi-\tfrac{r}{2}\Phi_{,r}-\tfrac{r}{2}\Psi_{,r}\;,\nonumber\\
&R_{\phi\phi}=\sin^2\theta R_{\theta\theta}\;,\nonumber
\end{align}\end{subequations}
}\vspace{2mm}\\
where $(...)_{,}$ denotes ordinary differentiation. The system above is solved by
\vspace{-2mm}\\
\Split{\begin{align}
&R=Ar^{\tfrac{2}{n-2}}\;,\nonumber\\
&\Phi=\sigma+\tfrac{\delta}{r}+\Phi_n\left(\tfrac{r}{L_q}\right)^{2\;\tfrac{n-1}{n-2}}\;,\nonumber\\
&\Psi=\tfrac{\delta}{r}+\Psi_n\left(\tfrac{r}{L_q}\right)^{2\;\tfrac{n-1}{n-2}}\;,\nonumber
\end{align}
}{
\begin{subequations}\label{pns}
\begin{align}
&A=\left[-\tfrac{6nq(3n-4)(n-1)}{(n-2)^2}\right]^{\tfrac{1}{2-n}}\;,\label{pns1}\\
&\Phi_n\equiv\!\!\left[-\tfrac{6n(3n-4)(n-1)}{(n-2)^2}\right]^{\tfrac{1}{2-n}}\!\!\!\tfrac{(n-2)^2}{6(3n-4)(n-1)}\;,\label{pns2}\\
&\Psi_n\equiv\!\!\left[-\tfrac{6n(3n-4)(n-1)}{(n-2)^2}\right]^{\tfrac{1}{2-n}}\tfrac{(n-2)}{3(3n-4)}\;,\label{pns3}
\end{align}
\end{subequations}
}\vspace{2mm}\\
where the integration constant $\delta$ has the dimensions of $[L]$ and the dimensionless integration constant $\sigma$ can be set equal to zero without loss of generality. The integration constant $A$ has the dimensions of $[L]^{(2n-2)/(2-n)}$, and $\Phi_n$ and $\Psi_n$ are dimensionless, accordingly. Moreover, one can check that $\Phi_n$ and $\Psi_n$ are well-defined only in the case $n=(2m+1)/\ell$ while we get $A>0$ since we assume $q<0$. In agreement to the geodesic motion as expanded in the weak field limit, the integration constant $\delta$ results equal to $\delta=-r^{S}$, where $r^S = 2GM$ is the \emph{Schwarzschild radius} of a central object of mass $M$.

The most suitable arena to evaluate the reliability and the validity range of the weak-field solution \reff{pns} is the Solar System \cite{zac, tak}. To this end, we can specify eqs.\reff{pns2}-\reff{pns3} for the typical length scales involved in the problem and we split $\Phi$ and $\Psi$ into two terms, the Newtonian part and a modification, \emph{i.e.},
\begin{subequations}
\begin{align}
&\Phi\equiv\Phi_N+\Phi_{M}\equiv-\rs/r+ \Phi_n(r/L_q)^{2(n-1)/(n-2)}\;,\label{lllqqlqlqlqlqllq}\\ 
&\Psi\equiv\Psi_N+\Psi_{M}\equiv-\rs/r+ \Psi_n(r/L_q)^{2(n-1)/(n-2)}\;,
\end{align}
\end{subequations}
here, the integration constant $\delta$ of eqs.\reff{pns2}-\reff{pns3} is $\delta=-\rs\equiv2GM_\odot$ ($M_\odot$ being the Solar mass). While the weak-field approximation of the Schwarzschild metric is valid within the range $\rs\ll r<\infty$ because it is asymptotically flat, the modification terms have the peculiar feature to diverge for $r\rightarrow\infty$. It is therefore necessary to establish a validity range, \emph{i.e.}, $r_{Min}\ll r\ll r_{Max}$, related to $n$ and $L_q$, where this solution is physically predictive \cite{extmetric}.

Since we aim to provide a physical picture at least of the planetary region of the Solar System, we are led to require that $\Phi_{M}$ and $\Psi_{M}$ remain small perturbations with respect to $\Phi_N$ and $\Psi_N$, so that it is easy to recognize the absence of a minimal radius except for the condition $r\gg \rs$. The typical distance $\Ls$ corresponds to the request 
\begin{equation}\label{rmax}
|\Phi_N(\Ls)|\sim|\Phi_{M}(\Ls)|\;,\qq\qq 
|\Psi_N(\Ls)|\sim|\Psi_{M}(\Ls)|\;.
\end{equation}
For $\rs\ll r\ll \Ls$, the system obeys thus Newtonian physics and experiences the post-Newtonian term as a correction. Another maximum distance $\Lss$ can be defined, according to the request that the weak-field expansion (\ref{line}) should hold, regardless to the ratios $\Phi_{M}/\Phi_N$ and $\Psi_{M}/\Psi_N$. $\Lss$ results to be defined by
\begin{equation}\label{star}
|\Phi_N(\Lss)|\ll|\Phi_{M}(\Lss)|\sim1\;,\qq\qq |\Psi_N(\Lss)|\ll|\Psi_{M}(\Lss)|\sim1\;.
\end{equation}
We remark that $\Ls$ and $\Lss$ are defined as functions of $n$ and $L_q$, \emph{i.e.},
\begin{equation}\label{LsLss}
\Ls\sim|\rs/\Phi_n|^{\tfrac{n-2}{3n-4}}L_q^{\tfrac{2n-2}{3n-4}}\;,\qq\qq
\Lss\sim L_q\big/|\Phi_n|^{\tfrac{n-2}{2n-2}}\;,
\end{equation}
and it is important to underline that, for the validity of our scheme, the condition $\Ls\gg\rs$ must hold, \emph{i.e.}, $L_q\gg\rs|\Phi_n|^{(n-2)/(2n-2)}$.  

Neglecting the lower-order effects concerning the eccentricity of the planetary orbit, we can deal with the simple model of a planet moving on circular orbit around the Sun with an orbital period $T$ given by $T=2\pi(r/a)^{1/2}$ ($a=(d\Phi/dr)/2$ being the centripetal acceleration). For our model, from eqs.(\ref{pns2}), we get
\begin{equation}\label{tibeta}
T_n=\frac{2\pi r^{3/2}}{(GM_\odot)^{1/2}}
\Big[1+2\Phi_n\;\frac{n-1}{n-2}
\Big(r^{\tfrac{3n-4}{n-2}}\Big)/\Big(\rs L_q^{\tfrac{2n-2}{n-2}}\Big)\Big]^{-1/2}\;.
\end{equation} 
We now can compare the correction to the Keplerian period $T_K=2\pi r^{3/2}(GM_\odot)^{-1/2}$, with the experimental data of the period $T_{exp}$ and its uncertainty $\delta T_{exp}$. We then impose the correction to be smaller than the experimental uncertainty, \emph{i.e.},
\begin{equation}\label{last}
\frac{\delta T_{exp}}{T_{exp}}\geqslant\frac{| T_K-T_n|}{T_K}\sim
|\Phi_n|\;\frac{n-1}{n-2}
\Big(r_P^{\tfrac{3n-4}{n-2}}\Big)/\Big(\rs L_q^{\tfrac{2n-2}{n-2}}\Big)\;,
\end{equation}
where $r_P$ is the mean orbital distance of a given planet from the Sun.
\newcommand{\pc}{\textrm{ pc}}

Let us now specify our analysis for the example of the Earth \cite{zac}. In this particular case, $T_{exp} \simeq 365.2563 \textrm{ days}$ and $\delta T_{exp}\simeq 5.0\cdot10^{-10} \textrm{ days}$ (with $r_P\simeq 4.8482\times10^{-6} \pc$). This way, for the Earth, we can get a lower bound $L_q>L^{Min}_{q\oplus}$ for the characteristic length scale of our model, as function of $n$, \emph{i.e.},
\begin{equation}\label{LMin}
L^{Min}_{q\oplus}(n)=\Big[1.3689\times10^{12}\;\frac{|\Phi_n|}{\rs}\;\frac{n-1}{n-2}\;
r_P^{\tfrac{3n-4}{n-2}}\Big]^{\tfrac{n-2}{2n-2}}\;,
\end{equation}
where $\Phi_n$ is defined in eq.(\ref{pns2}) and $L^{Min}_{q\oplus}\sim4\times10^{-3} \pc$, for a typical value $n\simeq2.66$. We remark that $L^{Min}_{q\oplus}$, by virtue of eq.(\ref{pns2}), is defined only for $n=(2m+1)/\ell$.

Our analysis clarifies how the predictions of the corresponding equations for the weak-field limit appear viable in view of the constraints arising from the Solar-System physics. Indeed, the lower bound for $L_q$ does not represent a serious shortcoming of the model, as we are going to discuss in Sec.\ref{sec:PR}, where a plot of $L^{Min}_{q\oplus}(n)$ and of $\Ls(L_q)$ and $\Lss(L_q)$ will be also addressed.

\section{Cosmological implementation of the f(R) model}

In order to study how our \emph{f(R)} model affects the cosmological evolution, we start from the modified gravitational action \reff{nonan} and we assume the standard Robertson-Walker (RW) line element in the synchronous reference system, \emph{i.e.},
\begin{align}
ds^{2}=dt^{2}-
a(t)^{2}\big[\,dr^{2}/(1-Kr^{2})+r^{2}d\Omega^{2}\,\big]\;,
\end{align}
where $a(t)$ is the scale factor and $K$ the spatial curvature constant. Using such expression, the 00-component of eq.\reff{variational} results, for symmetry using the Bianchi identity, the only independent one and it writes as
\begin{align}\label{eq-00}
f^{\pr}R_{00}-\tfrac{1}{2}\,f+3(\ad/a)\,f^{\pr\pr}\,\dot{R}=\chi\,T_{00}\;.
\end{align}
where the dot indicates the time derivative. We assume as matter source a perfect-fluid EMT, \emph{i.e.}, $T_{\mu\nu}=(p+\rho)\,u_\mu u_\nu-pg_{\mu\nu}$, in a comoving reference system (thus $T_{00}=\rho$), where $p$ is the thermostatic pressure, $\rho$ the energy density and $u_\mu$ denotes the 4-velocity. The 0-component of the  conservation law, \emph{i.e.}, $T_{\mu;\,\nu}^{\nu}=0$ with $\nu=0$, assuming the \emph{equation of state} (EoS) $p=w\rho$, gives the following expression for the energy density: $\rho=\rhoo\,[a/\azero]^{-3(1+w)}$.

Using now $f=R+qR^n$ with $q<0$, we are able to explicitly write eq.\reff{eq-00}:
\begin{align}\label{a-eq}
&2\,\tilde{\chi}\,a^{1-3w} \;+\;
6^{n}\,n\,q\,a^{5-2n}\,\ddot{a}\,(-K-\dot{a}^2-a\,\ddot{a})^{n-1}
\;+\;\;\;\nonumber\\
&+a^2\,
\big[-6K-6\,\dot{a}^2+6^n\,q\,a^{2(1-n)}\,(-K-\dot{a}^2-a\,\ddot{a})^{n}\big]
\;+\;\\
&+6^n(n-1)n\,q\,\dot{a}\,a^{2(2-n)}\,(-K-\dot{a}^2-a\,\ddot{a})^{n-2}
\big[-2\,\dot{a}^3-2K\,\dot{a}+a\,\dot{a}\,\ddot{a}\,+a^2\,\dddot{a}\big]=0\;,\nonumber
\end{align}
where $\tilde{\chi}=\chi\rhoo\azero^{3(1+w)}$. Let us now assume a power-law $a=\azero\;[t/\tzero]^x$ for the scale factor and, for the sake of simplicity, we set $\ao=\azero^{\phantom x}\tzero^{-x}$ (clearly, $[\ao]=[L^{1-x}]$). Here and in the following, we use the subscript $(...)_{\scriptscriptstyle 0}$ to denote quantities measured today. In this case, eq.\reff{a-eq} can be recast in the form
\begin{align}
&-6\ao^{2} K\, t^{2x}-6\ao^{4}x^{2}\,t^{4x-2}+
q\ao^{4}\,t^{4x}\,\Big(C_1\,t^{-2}-6\ao^{-2} K\, t^{-2x}\Big)^{n}+
\;2\tilde{\chi}\ao^{1-3w}\,t^{x(1-3w)}\;=\;\nonumber\\
&=\;nqx\ao^{6-2n}\,t^{6x}\,\Big(C_1\ao^{2}\,t^{-2}-6K\,t^{-2x}\Big)^{n}\;
\frac{(C_2 K\,t^2 + xC_3\,t^{2x})}{(K\,t^2 + C_4\,t^{2x})^{2}}\;,
\label{power-law-eq}
\end{align}
where $C_1=6x(1-2x)$, $C_2=(x(2n-1)-1)$, $C_3=x\ao^{2}(x+2n-3)(2x-1)$, and \\$C_4=x\ao^{2}(2x-1)$.

\subsection{Radiation-dominated Universe}
Here, we assume the radiation-dominated Universe EoS $w=1/3$ ($\rho\sim a^{-4}$). In the following, we will discuss the three distinct regimes, in the asymptotic limit as $t\to 0$, for $x<1$, $x>1$ and $x=1$, separately.

In the case $x<1$, all terms containing explicitly the curvature $K$ of eq.\reff{power-law-eq} results to be negligible for $t\to0$ and asymptotic solutions are allowed if and only if $x\leqslant n/2$ which, in the case $2<n<3$ we are considering, is always satisfied. The leading-order term of eq.\reff{power-law-eq} writes as
\begin{align}\label{radiation-xmin1}
q\ao^{4}C_1^{n}\;[\,1-\,(C_3/C_4^2)nx^2\ao^2\,]\;t^{4x-2n}=0\,,
\end{align}
and $x=1/2$ and $x=[2+3n-2n^2\pm(4+8n+n^2-12n^3+4n^4)^{1/2}]/2n$ are the solutions. Such second expression results to be negative or imaginary for $2<n<3$ and must be excluded. Thus, the only solution for $x<1$, in the asymptotic limit for $t\to0$, is the well-known radiation dominated behavior $a\sim t^{1/2}$. In the other two cases, \emph{i.e.}, for $x\geqslant1$, it is easy to recognize that no asymptotic solutions are allowed. Therefore, the approach to the initial singularity is not characterized by power-law inflation behavior when spatial curvature is non-vanishing.

Let us now assume a vanishing spatial curvature in eq.\reff{power-law-eq}. In can be show how, for $K=0$, the radiation-dominated solution with $w=1/3$ and $x=1/2$ is an \emph{exact} solution (non-asymptotic and allowed for all $n$-values) giving $\rhoo=3/(4\chi\tzero^{2})$, matching the standard FLRW case. In the case $x>1$, the leading-order terms of eq.\reff{power-law-eq} read, for $t\to0$ and $K=0$,
\begin{align}\label{radiation-xmagg1}
q\ao^{4}C_1^{n}\;[\,1-\,(C_3/C_4^2)nx^2\ao^2\,]\;t^{4x-2n}\;+\;2\tilde{\chi}=0\;.
\end{align}
Three distinct regimes have to be now separately discussed. For $x>n/2$, the leading order of the equation above does not admit solutions since it writes simply $2\tilde{\chi}=0$ and, for $x<n/2$, the solutions of eq.\reff{radiation-xmagg1} are those obtained in the case for $x<1$. Instead, for $x=n/2$, and defining $\Hzero=(n/2)/\tzero$, one gets
\newcommand{\qo}{q_{\scriptscriptstyle 0}}
\begin{align}\label{rhozerooooo}
\rhoo=\frac{\tilde{\,\,\rhoo}(n)\,\qo}{4\chi\tzero^{2}}\;,\;
\tilde{\,\,\rhoo}(n)=\tfrac{\,\,\,3^{n}}{2}\;(1-n)^{(n-1)}n^n(n(4+(6-5n)n)-4)(n/2)^{2-2n},
\end{align}
where we have introduced the dimensionless parameter $\qo^{\phantom2}=\Hzero^{2n-2}\,q$. We remark that the constraint $n=(2m+1)/\ell$ (which is in agreement with respect to the one obtained from Solar-System test) must hold in order to have $\rhoo>0$ since we have assumed $q<0$ and therefore $\qo<0$. The function $\tilde{\,\,\rhoo}$ results to increase as $n$ goes from $2$ to $3$ and, in particular, one can get $216<\tilde{\,\,\rhoo}<21\;024$.  Finally, for $x=1$ and $K=0$, eq.\reff{power-law-eq} reads $[1-n(2n-2)]\;t^{4-2n}=0$, giving $n=[1\pm\sqrt{3}]/2$. As the previous case, the regime $x=1$ does not admit solutions in the region $2<n<3$.

\subsection{Matter-dominated Universe}
Let us now study the matter-dominated Universe EoS $w=0$ ($\rho\sim a^{-3}$). As previously done, we analyze the three distinct regimes for $x<1$, $x>1$ and $x=1$, and, in the limit for $t\to\infty$, it is easy to recognize that there are no power-law solutions in all these cases for $K\neq0$. Setting $K=0$, the $x\geqslant1$ regimes do not provide any power-law form for cosmological dynamics either. On the other hand, for $x<1$ and assuming zero spatial curvature in eq.\reff{power-law-eq}, we get the following equation:
\begin{align}\label{matter-k0}
[-6x^{2}\ao^{4}]\;t^{4x-2}+2\tilde{\chi}\ao\;t^{x}=
\ao^{4}qC_1^{n}[-1+C_3\ao^{2}nx^{2}/C_4^{2}]\;t^{4x-2n}\;.
\end{align}
Since $4x-2>4x-2n$, the term on the right hand side can be neglected in the limit of large $t$ and the equation above admits three distinct situations: $x<2/3$, $x>2/3$ and $x=2/3$. Both cases with $x\neq2/3$ do not admit solution. The case $x=2/3$ admits instead an asymptotic solution for $t\to\infty$. In fact, eq.\reff{matter-k0} reduces to $8\ao^{3}=6\tilde{\chi}$ and the FLWR matter-dominated power-law solution $a=\ao\,t^{2/3}$ is reached setting $\rhoo=4/(3\chi\tzero^{2})$.

In conclusion, we can infer that, for $f(R)=R+qR^n$, the standard matter-dominated FLRW behavior of the scale factor $a\sim t^{2/3}$ is the only asymptotic (as $t\to\infty$) power-law solution.

\subsubsection{Range of t-values:}\q As shown above, the matter dominated solution $a\sim t^{2/3}$ is obtained for $K=0$ and asimptotically as $t\to\infty$. In order to neglect all the $K$-terms in our $f(R)$ model, we start directly from the expression of the Ricci scalar \cite{Kolb}. Using a power-law scale factor, we get the $t$-range (if $x\neq1/2$ and $x<1$)
\begin{align}\label{t-contraint}
t \ll
\big|\;[x(2x-1)]/[K/\ao^{2}]\;\big|^{1/(2-2x)}\;.
\end{align}
For the matter-dominated era and using standard cosmological parameters \cite{WMAP}, one can get the upper limit $K/\ao^2\;{\scriptstyle \lesssim}\; 0.006\; (\Hzero)^{2/3}$, to estimate the value of $K/\ao^{2}$. Thus, setting $x=2/3$, we get the bound $t\ll\;235/\Hzero$, independently of the form of $f(R)$.

At the same time, if we set $x=2/3$, the asymptotic solution $\rhoo=4/(3\chi\tzero^{2})$ is reached neglecting the right hand side ($\ll1$) of eq.\reff{matter-k0}, \emph{i.e}, if $t$ is constrained by the following lower limit (we remind that $\qo^{\phantom2}=\Hzero^{2n-2}\,q$)
\begin{align}
t\gg\;\mu(n,\qo)/\Hzero\;,\qq
\mu(n,\qo)=\big|\;\qo\big[-(4/3)^n + 2^{(2n+1)}3^{-n}\,n(2n-7/3)\big]\;\big|^{1/2(n-1)}\;.
\label{l_tmin}
\end{align}\newcommand{\secondi}{\textrm{s}}
Let us now recall that the matter-dominated era began, assuming $\Hzero^{-1}\simeq4.3\times10^{17} \secondi$, at $t_{Eq}\simeq5.1\times10^{-6}/\Hzero$. In this sense, we can safely assume $\mu(n,\qo)\leqslant5.1\times10^{-8}$, which implies an upper limit for $|\qo|$, $i.e.$, $|\qo|\leqslant|\qo|^{Max}$, where
\begin{align}\label{q0maxFORMULA}
|\qo|^{Max}(n)=\big[5.1\times10^{-8}\big]^{2(1-n)}\;\big|-(4/3)^n + 2^{(2n+1)}3^{-n}\,n(2n-7/3)\big|^{-1}\;.
\end{align}
It is easy to check that the function $|\qo|^{Max}(n)$ is decreasing as $n$ goes from $2$ to $3$, in particular, one gets: $10^{-16}\;{\scriptstyle \lesssim}\;|\qo|^{Max}\;{\scriptstyle \lesssim}\;10^{-31}$.

\section{The inflationary paradigm}

After discussing the power-law evolution of the Universe proper of the radiation- and matter-dominated eras, we now analyze the inflationary behavior characterizing the very early dynamics (for an interesting approach to the inflationary scenario within the modified gravity scheme, see \cite{Odintsov1, Odintsov2}). In this respect, we hypothesize an exponential behavior for the scale factor of the Universe $a=\azero\;e^{s(t-\tzero)}=\ao\;e^{st}$, where $s>0$ and $\ao=\azero e^{-s\tzero}$. In the following, we concentrate the attention on the solution for vanishing spatial curvature $K=0$ and, in this case, eq.\reff{a-eq} rewrites as
\begin{align}
\ao^{4}\;e^{4st}\big[q(-12)^n\;s^{2n}(1-n/2)-6s^{2}\big]+2\tilde{\chi}(\ao\,e^{st})^{1-3w}=0\;.	
\end{align}
Let us now assume $w=-1$ (\emph{i.e.}, $\rho=\rho_I=const.$) during inflation. Using the definition $\qo^{\phantom2}=\Hzero^{2n-2}\,q$, the equation above reduces to
\begin{align}\label{eq:inflation}
\big[(-1)^{n}12^n\;\qo(1-n/2)\big]\so^{2n}-6\so^{2}+\kappa=0\;,
\end{align}
where $\kappa=2\chi\rho_I\Hzero^{-2}$ and $\so$ is a dimensionless parameter defined as $\so=s/\Hzero$. Since $\Hzero$ denotes the Hubble parameter measured today and estimating $H_I=\sqrt{\chi\rho_I/3}$ (\emph{i.e.}, accordingly to its Friedmannian value) during inflation \cite{Kolb}, one can obtain $\kappa\sim H_I^{2}/\Hzero^{2}\sim10^{100}$. For such values, it is easy to realize that considering the case $n=2\ell/(2m+1)$, the equation above does not admit real solution, thus we now discuss, consistently with the previous analyses, only $n=(2m+1)/(2\ell+1)$.

In order to integrate eq.\reff{eq:inflation}, we focus on a particular value of the power-law $f(R)$ exponent, \emph{e.g}, $n=29/13\sim2.23$. Using eq.\reff{LMin}, for this value of $n$ one obtains that it can be safely considered $L^{Min}_{q\oplus}\sim1.44\times10^{-5}\pc$ and, having in mind that $L_q=|\qo|^{1/(2n-2)}/\Hzero$ with $\Hzero^{-1}\simeq4.2\times10^{9} \pc$, we get $|\qo|>|\qo|^{Min}\sim2.56\times10^{-36}$. 

Let us now fix the parameter $\qo$ to a reasonable value like $\qo^{*}\sim-10^{3}|\qo|^{Min}$ (such assumption will be physically motivated in the next Section). In this case, the solution of eq.\reff{eq:inflation} is $\so\sim2.45\times10^{29}$. This analysis demonstrates that an exponential early expansion of the Universe is still associated to a vacuum constant energy, even for the modified Friedmann dynamics. However, we see that the rate of expansion is significantly lower than the Friedmann-like one of about a factor in $\so$ of $10^{20}$. Although our estimation relies on the Friedmannian relation between $H_I$ and $\rho_I$ (the latter is taken of the order of the Grand Unification energy-scale), nevertheless the values of $\so$ remains many order of magnitude below the standard value $\sim10^{50}$ even if we change $H_I$ for several order of magnitude. Despite this difference, it is still possible to arrange the cosmological parameter in order to have a satisfactory inflationary scenario, as far as we require a longer duration of the de Sitter phase.

\section{Physical remarks}\label{sec:PR}

\begin{figure}[!ht]
\includegraphics[width=0.5\hsize]{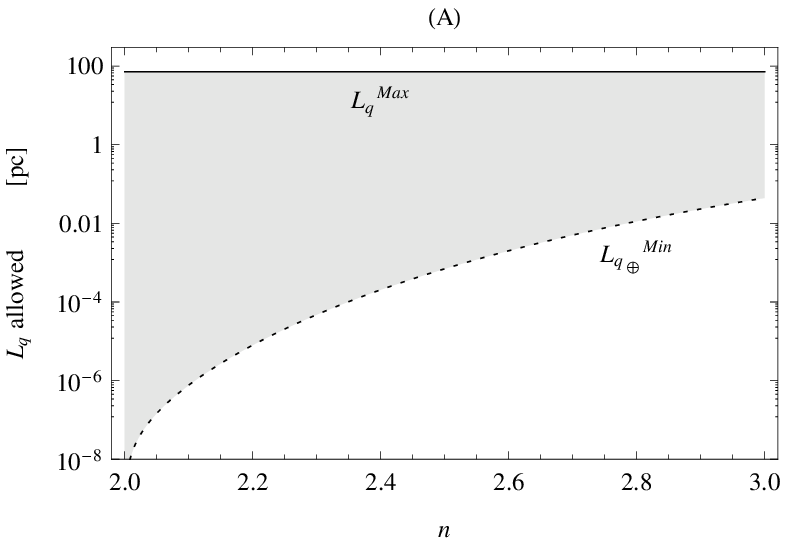}
\includegraphics[width=0.5\hsize]{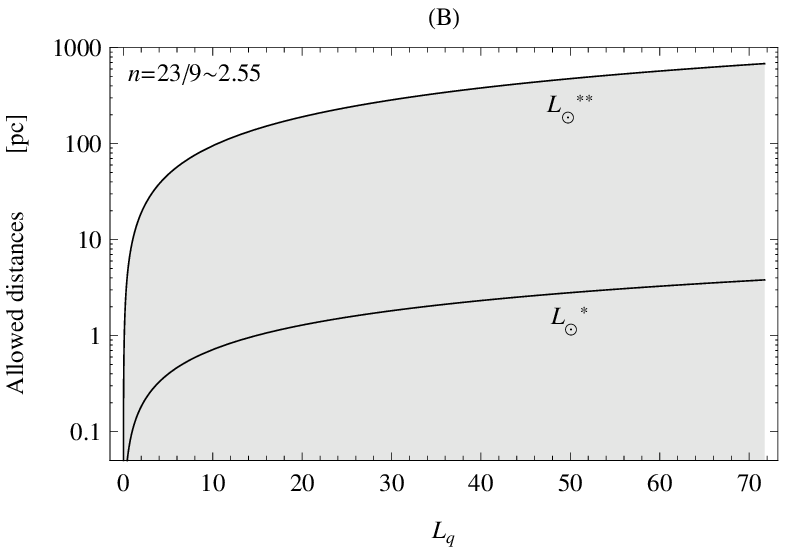}
\caption{\textbf{Panel A:} $L_{q\oplus}^{Min}$ of eq.\reff{LMin} and $L_q^{Max}$ of eq.\reff{LMax....}. The gray zone represents the allowed characteristic-length scales of the model. We stress that $L_{q\odot}^{Min}$ is defined only if $n=(2m+1)/\ell$, as represented by the dotted line. \textbf{Panel B:} $L_{\odot}^{*}$ and $L_{\odot}^{**}$ of eq.\reff{LsLss}. The gray zone represents here the allowed distances for the model.}
\label{length}
\end{figure}

As already discussed in Sec.2, the parameter $q$ has dimension $[L]^{2n-2}$. We have therefore defined a characteristic length scale of the model as $L_q(n)=|q|^{1/(2n-2)}$. Assuming $f(R)$ corrections to be smaller than the experimental uncertainty of the orbital period of the Earth around the Sun, the lower bound \reff{LMin} for $L_q(n)$ was found. In order to identify the allowed scales for our model and in view of the upper constraint on the parameter $\qo^{\phantom2}=\Hzero^{2n-2}\,q$ derived in the cosmological framework, we can now define the upper limit for $L_q(n)$ as 
\begin{align}\label{LMax....}
L_q^{Max}(n)=[|\qo|^{Max}]^{1/(2n-2)}/\Hzero\;,
\end{align}
which, considering eq.\reff{q0maxFORMULA}, yields to the constraints $65.59 \pc<L_q^{Max}<78.37 \pc$, for $2<n<3$. Assuming $\Hzero^{-1}\simeq4.2\times10^{9} \pc$, the two bounds for the characteristic length scales here discussed, \emph{i.e}, eq.\reff{LMin} and eq.\reff{LMax....}, are plotted in Fig.\ref{length}(A).
At the same time two other typical lengths have been outlined in eq.\reff{LsLss} for the Solar System. $L_{\odot}^{*}$ represents the minimum distance to have post-Newtonian and Newtonian terms of the same order. While $L_{\odot}^{**}$ was defined according to the request that the weak-field expansion holds. Setting now $n=23/9\simeq 2.55$, one can show from eq.\reff{LMin} and eq.\reff{LMax....} that the allowed scales are $0.0013 \pc\;{\scriptstyle \lesssim}\;L_q\;{\scriptstyle \lesssim}\;71.72 \pc$. In this range, $L_{\odot}^{*}$ and $L_{\odot}^{**}$ can be plotted as in Fig.\ref{length}(B).

Summarizing, our analysis states a precise range of validity for the power-law $f(R)$ model we consider. Indeed, for a generic value of $n$ (\emph{i.e.}, not close to $2$ or $3$) the fundamental length of the model is constrained to range from the super Solar-System scale up to a sub-galactic one. Therefore, in agreement to eq.\reff{lllqqlqlqlqlqllq}, we have to search significant modification for the Newton law in gravitational system lying in this interval of length scales, like for instance, stellar clusters. 

{\small\textbf{Acknowledgment}: NC gratefully acknowledges the CPT - Universit\'e de la Mediterran\'ee Aix-Marseille 2 and the financial support from ``Sapienza'' University of Rome.}

\end{document}